\documentclass[runningheads]{llncs}
\renewcommand{\and}{\\}
\usepackage[dvips]{graphicx}
\usepackage{amsmath}
\usepackage{color}
\usepackage{placeins}
\usepackage[normalem]{ulem}

\begin{document}

\newcommand{\JJ}{\text{JJ}}
\newcommand{\IJbb}{I_J^{B}}
\newcommand{\CJbb}{C_J^{B}}
\newcommand{\IJab}{\hat I_J}
\newcommand{\CJab}{\hat C_J}
\newcommand{\Lhat}{\hat L}

%
\title{Ballistic reversible gates matched to bit storage:\newline
Plans for an efficient CNOT gate using fluxons}
\titlerunning{Ballistic reversible fluxon gates matched to bit storage}
%
\author{Kevin D. Osborn\inst{1,2} and
Waltraut Wustmann\inst{1}}
\authorrunning{K.D. Osborn et al.}
%
\institute{
Lab. for Physical Sciences, University of Maryland, College Park, MD 20740, USA \and
Joint Quantum Institute, University of Maryland, College Park, MD 20742, USA
}
\maketitle              
\begin{abstract}
New computing technologies are being sought near the end of CMOS transistor scaling, meanwhile superconducting digital, i.e., single-flux quantum (SFQ), logic allows incredibly efficient gates which are relevant to the impending transition. In this work we present a proposed reversible logic, including gate simulations and schematics under the name of Reversible Fluxon Logic (RFL).
In the widest sense it is related to SFQ-logic, however it relies on (some approximately) reversible gate dynamics and promises higher efficiency than conventional SFQ which is logically irreversible.
Our gates use fluxons, a type of SFQ which has topological-particle characteristics in an undamped Long Josephson junction (LJJ).
The collective dynamics of the component Josephson junctions (JJs) enable ballistic fluxon motion within LJJs as well as good energy preservation of the fluxon for JJ-circuit gates.
For state changes, the gates induce switching of fluxon polarity during resonant scattering at an interface between different LJJs.
Related to the ballistic nature of fluxons in LJJ, the gates are powered, almost ideally, only by data fluxon momentum in stark contrast to conventionally damped logic gates which are powered continuously with a bias. At first the fundamental Identity and NOT gates are introduced.
Then 2-bit gates are discussed, including the IDSN gate which actually allows low fluxon-number inputs for more than 4 input states.
A digital CNOT, an important milestone for 2-bit reversible superconducting gates, is planned as a central result.
It uses a store and launch gate to stop and then later route a fluxon. This use of the store and launch gate allows a clocked CNOT gate and synchronization within.
The digital CNOT gate could enable high efficiency relative to conventional irreversible gates and shows the utility of the IDSN as a reversible gate primitive.
\end{abstract}
%
\section{Introduction}
Now, with the end of industry-standard CMOS scaling in sight,
diverse technologies are being sought for an impending technology transition,
including ones that lower on-chip dissipation from logic gate power \cite{IRDS2018}.
Superconducting computing, i.e., Single-Flux Quanta (SFQ) logic,
offers two paths to greater efficiency in digital logic \cite{Holmes2013,Soloviev2017}.
The first and prevalent one uses conventional SFQ gates, where static power
has been greatly reduced and practically eliminated over the last five years
\cite{Likharev1991,HerrETAL2011,Mukhanov2011}.
These gates are logically irreversible as are the ones in CMOS,
meaning that a gate input cannot be reconstructed from its output.
A second, more dramatic, path to efficiency adds physical reversibility in digital logic.
Theoretically an ideal reversible computation has no minimum in energy cost,
partially through the avoidance of bit erasure \cite{Landauer1961,Bennett1973}.
Reversible superconducting digital logic types have been demonstrated \cite{QFPgate,RenSem2011} with record breaking efficiency. While current superconducting quantum processors, which may use inefficient control fields,
show that precise state reversibility is possible,
reversible digital logic provides another dramatic opportunity
--- using physical reversibility for digital-computing energy efficiency, including all control fields.

Rolf Landauer studied the minimum thermodynamic dissipation for bit erasure,
$\ln(2) k T = 0.69 kT$,
where $k$ is the Boltzmann constant and $T$ is the temperature,
and accordingly we anticipate this as the minimum energy dissipation
in every irreversible logic gate \cite{MarNorVed2009}.
In one typical such gate, two bits of input are used to create one bit of output,
e.g., a NAND or XOR gate.
Switching (including bit changes) in conventional gates is performed using potential (stored) energy,
where at least this amount is dissipated during the process.
In CMOS technology a potential energy of $CV^2/2$ is dissipated, where $C$ is the gate capacitance and $V$ is the supplied voltage.
As a result, operations with large capacitance (e.g. fan-out, or long connections) are especially inefficient.
Dissipation of conventional SFQ (C-SFQ) gates is generally understood
from the model of a current-biased Josephson junction with damping \cite{Tolpygo2016}.
In C-SFQ, Josephson junctions (JJs) generally change phase (or switch) by $\approx 2\pi$, while an SFQ moves as part of the gate operation.
For example, an SFQ at particular location and time represents the 1-state,
and its absence represents the 0-state.
For typical bias which is near the critical current $I_c$ of a JJ, each switching (or movement) typically dissipates an energy close to $E_{\JJ\,\text{switch}} \approx I_c \Phi_0$,  where $\Phi_0$ is the magnetic flux quantum.
Incidentally, SFQ generally dissipates a similar energy for motion between cells in a Josephson transmission line (JTL), a structure that is useful for connecting gates over short distances.
One type of C-SFQ logic has operated at a low critical current to allow a switching energy
corresponding to $I_c \Phi_0 \approx 1300 k T$ \cite{HerrETAL2011,PrivComm2017}.

One way to understand the energy limits of conventional gates
is to first look at the minimum potential energy barrier $U_B = \gamma_T kT$ with $\gamma_T \sim 10$
which separates two meta-stable states with sufficient thermal stability:
the current state and the next possible state of the logic cell.
In C-SFQ logic this barrier is determined by the intrinsic JJ barrier between wells
$U_{\JJ} \approx I_c \Phi_0/\pi$, reduced by a factor of $4$ due to an applied current bias
to give the energy barrier
$U_B \approx I_c \Phi_0/4\pi \approx E_{\JJ\,\text{switch}}/4\pi$.
One can therefore generally expect that JJs in conventional SFQ gates have
an energy for switching of $E_{\JJ\,\text{switch}} \approx 4\pi U_B \approx 4\pi \gamma_T k T$.
Furthermore, gate dynamics evolves in a multi-step process generally involving $n \sim 8$
switching JJs (and SFQ movements) \cite{Tolpygo2016},
resulting in a gate energy of
$\approx 4\pi n \gamma_T k T \approx n 100 k T \approx 1000 k T$.
Complications such as the presence of a third meta-stable state and fabrication variations
can increase the practical energy cost further.
Therefore, in this estimate, conventional logic is expected to dissipate $>1000 k T$ per gate operation.
These estimates demonstrate that conventional (superconducting) logic operates orders of magnitude above the bit-erasure limit ($\ln(2) kT$),
and the energy is rather limited by the requirement of thermal state stability
and the damped dynamics used in conventional state switching
(i.e., $\gamma_T kT$ multiplied by additional factors).
Physical reversibility in logic can allow energy costs per gate below the bit erasure limit.
However, reversible logic gates with state-of-the-art efficiency can be of technological utility as well as scientific interest.
For this latter pursuit in physical reversibility,
we define Reversibly-Enabled Energy Efficiency (REEF)
as a property of a logic function which beats state-of-the-art conventional
gate dissipation: $E_{\text{REEF}\,\text{op}} < 1000 k T$.
\\

Adiabatic reversible logic allows reversible physical state evolution
to target notable goals such as achieving REEF or a fundamental limit (e.g. the minimum bit erasure energy).
Here a slow (adiabatic) modulation of the potential-energy landscape with barrier
within the gate allows the physical state to evolve near the potential minimum
over a long time compared to the characteristic damping time \cite{Lik1982}.
The state dynamics are approximately reversible in time through reversal of the modulation. Though it has an inherent speed limitation (maximum modulation rate),
superconducting circuits nevertheless enable practical clock (modulation) frequencies. Often a small JJ oscillation period is used with a longer damping time. This property makes the circuit underdamped, in contrast to C-SFQ.
In an adiabatic logic with N-SQUIDs, a shift register was demonstrated \cite{RenSem2011}
with dissipation reported at a few $kT$.
Adiabatic Quantum Flux Parametron (AQFP) logic is also known to allow regimes
of good physical reversibility (in logically reversible operations).
What is also interesting is that AQFP allows demonstrations with non-adiabatic dynamics,
in reversible or irreversible gates, where these abrupt dynamics are comparable
to the JJ-switching process in C-SFQ.
Dissipation was demonstrated to be very low in one demonstration \cite{QFPE}, where one AQFP cell was switched between its two states, and switching was measured at $10 \,\text{zJ} (10^{-20} \,\text{J})$ per switch
using a $5 \,\text{GHz}$ modulation clock. This dissipation is equal to $180 kT$  at $T=4 \,\text{K}$,
and is a demonstration of reversible state changes despite the fact that it is not yet optimized for efficiency. Simulations show that further optimization will allow the device to function below the bit erasure limit \cite{YoshikawaETAL2015}.
A reversible AQFP gate with three inputs and three majority gate functions made from many AQFP cells was demonstrated \cite{QFPgate}.
Similar to other adiabatic logic types, the simulations show that dissipated energy is inversely proportional to clock (modulation) period.

\section{Using a ballistic reversible approach}

Unlike the adiabatic type of reversible logic, where an external field drives gate evolution,
ballistic (or more generally, scattering-based) logic is made to utilize particle momenta
for energy-conserving gates \cite{FredTof1982}.
In the classical model, billiard balls ideally collide on a table (a two-dimensional space)
and the logic state follows from the absence or presence along paths produced by collisions.
Ballistic logic has been recently proposed with specific asynchronous gate functions \cite{Frank2017}.
The classic model, however, is synchronous and spatial and timing control is desired
to prevent errors from imperfections.
Furthermore, one may generally desire some synchronization for basic parallel computations and memory access.
In this work we propose the further use of dynamically reversible gates without dissipation
named Reversible Fluxon Logic (RFL) \cite{WusOsb2017}.
Furthermore, we will mate them to bit storage between gates,
with some intentional damping, for synchronization and clocking to create a CNOT with REEF efficiency.

Our approach will conserve the number of SFQ similar to the billiard balls in the classical approach.
However, our approach is also different than the classical model because we will use fixed paths for scattering, independent of the input states.
In a recent quantum application, a fluxon in underdamped long-Josephson junctions (LJJ)
was used for continuously powered qubit readout \cite{UstinovETAL2014}.
In our proposed logic, fluxons are used in undamped LJJs without bias for logic.
Same-energy fluxon and antifluxon states --- the two lowest-energy topological solitons
in the Sine-Gordon equation \cite{McLaughlinScott1978}
--- will be used for the 0 and 1 bit states, respectively.
By using the LJJ without bias we avoid the dissipation associated with accelerating the SFQ,
and the motion is unpowered unlike the switching of individual JJs in JTLs or C-SFQ.
Thus the starting point of the design avoids the energy dissipation
$E_{\JJ\,\text{switch}}$
associated with C-SFQ or any logic with a potential energy difference between states.

RFL is enabled by the collective dynamics of JJs, both in the LJJs and in JJ-circuit interfaces between LJJs.
An LJJ modeled with discrete components look schematically like a JTL.
However due to a small ratio of linear to Josephson inductance the fluxon in our LJJ is spread over multiple cells, while an SFQ in a JTL is approximately confined in one cell.
The fluxon is a topological soliton, according to the underlying Sine-Gordon equation, and allows it to move ballistically (with a fixed shape over a long distance relative to its size and without a large change in velocity).
The motion of such a flux-soliton thus is an energy-conserving process
where the undamped JJs switch by $2\pi$ in coordinated collective dynamics with their neighboring JJs.

Similarly we exploit collective JJ dynamics in the development of ballistic gates.
Our reversible gates are designed to change fluxon polarity (direction of magnetic flux)
from fluxon to antifluxon or vice versa for bit-state changes,
such that one JJ in a NOT gate will undergo a $4\pi$-phase change.
Previously studied gates include the NOT, Identity (ID), and the NSWAP=NOT(SWAP) \cite{WusOsb2017}.
The gate structure consists of input and output LJJs connected by a circuit interface
(where the choice of inputs and outputs could be exchanged due to dynamic reversibility).
The input fluxons enter LJJs in a gate and then excite oscillatory dynamics centered at the gate interface.
From those localized oscillations, fluxons emerge after a short time in the output LJJs, where the polarities of the output fluxons represent the result of the gate operation.
The gate dynamics contrast but also augment previously studied soliton phenomena.
The dynamics are currently understood from an analysis which assumes one quasiparticle in each LJJ consisting of a superposition of fluxon and mirror antifluxon. The quasiparticles of the gate are coupled through an effective mass matrix.
Analysis of these quasiparticle dynamics shows fluxon-number conserving scattering in the NOT and ID gates \cite{WusOsb2017}. The reversible dynamics are undamped, but the fluxons lose some kinetic energy in the process.
In some simulations the output fluxons have only a slightly diminished velocity relative to the input fluxons, showing that fluxon energy is nearly conserved (e.g. $>97\%$).
Even though the dynamics are not ideally reversible, operational gates conserve all the potential (rest) energy of fluxons and take the fluxons to subsequent stages.
Using tentative numbers such as $I_c \sim 100 \, \text{nA}$ for the parameters simulated, the fluxon energy in the logic becomes $\sim k \cdot 60 \,\text{K}$ or $\sim 20 kT$ for $T=3\,\text{K}$.
For $5\%$ energy loss, the gate energy might be as low (and efficient) as $E_{\text{REEF}\,\text{op}} \sim 1 kT$.
\\

We also introduce a similarly efficient 2-input RFL gate called the IDSN or ``IDeSaN'' gate,
so-named because it performs identity (ID) as a default for single-fluxon input, and for the input of two fluxons of the Same polarity it performs a NOT operation.
A reversible CNOT gate does not appear in previous superconducting SFQ logic,
but represents an important mathematical class as a 2-bit digital reversible gate
--- the non-degenerate linear affine class \cite{Aaronson2017}.
To design a CNOT gate, we couple IDSN and NOT gates with bit Storage aNd Launch (SNL) gates.
In these gates a fluxon will stop for storage until a clock pulse launches it into one of two LJJs as determined by the bit state (fluxon polarity).
This gate is triggered by a clock fluxon.
Moreover, the synchronous launch from two independent SNL gates can be implemented by splitting a clock fluxon.
They will also enable use of the IDSN by routing the fluxons for the input cases of the IDSN during the launch.
Despite the loss of dynamical reversibility in the SNL gate leading to an energy cost for stopping and relaunching fluxons, we provide promising results on the CNOT as a REEF gate.

\section{Fundamental gates in Reversible Fluxon Logic (RFL)}

\begin{figure}[ptb]\centering
 \includegraphics[width=\textwidth]{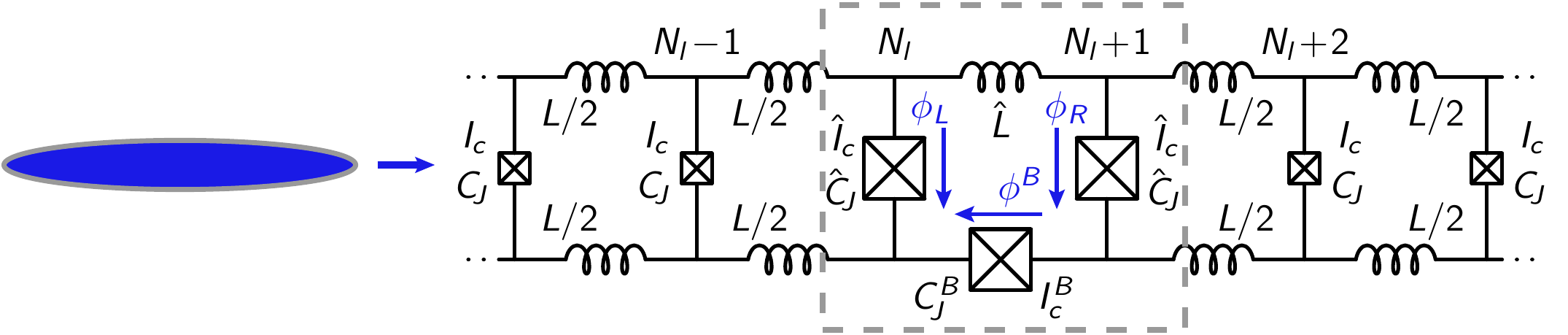}
\caption{
One-bit gate structure. The structures marked with ``X'' are Josephson junctions (JJs) without shunt resistors.
JJs are connected by superconducting wires of inductance $L/2$.
A fluxon is illustrated to enter the circuit,
where the Josephson penetration length is approximately three cells large and shown as the fluxon-core size (blue ellipse).
It may travel into the first LJJ, defined as repeated cells to the left of the dashed box. toward the interface, defined as circuit elements in the dashed box.
This excites short-lived oscillatory dynamics centered at the interface.
For a one-bit NOT gate, an antifluxon will then exit on the output LJJ (to the right of dashed box),
where fluxon polarity is inverted on the output relative to the input.
Example parameters for a NOT gate include interface parameters of $\CJab \approx 5.8 C_J$, $\IJab \approx 0.8 I_c$, and $\CJbb \approx 12 C_J$.
Negligible or small parameters are allowed elsewhere such as $\IJbb \approx 0.1 I_c$ and $\Lhat \approx 0.06 L$.
}
\label{fig1}
\end{figure}

In Figure ~\ref{fig1} we show a NOT gate schematic. A fluxon can approach from the left LJJ as illustrated.
In our simulations we typically use  $L_J = \Phi_0/(2\pi I_c) = 7 L$,
where the characteristic length of a static fluxon, or Josephson penetration length,
is $\lambda_J = a \sqrt{L_J/L} = 2.65 a$, where $a$ is the unit cell size.
The fluxon length decreases with speed but undesired discreteness effects are not present sufficiently below the maximum velocity
$c=\omega_J \lambda_J$, where $\omega_J = 1/\sqrt{L_J C_J}$
is the JJ frequency and $C_J$ is the capacitance of JJs in the LJJ.
It is helpful to note that even though the LJJ is modeled discretely and will have interfaces at the gate, its dynamics is well approximated by the continuous Sine-Gordon equation,
\begin{equation}
   \frac{\text{d}^2\,\phi}{\text{d}t^2}
  - c^2 \frac{\text{d}^2\,\phi}{\text{d}x^2}
  + \omega_J^2 \sin \phi = 0
  \,.
\end{equation}
In accord with the soliton solution to the Sine-Gordon equation,
at $0.6 c$, as we use below, the fluxon length is only decreased by $20\%$
such that ballistic motion between gate interfaces can be maintained with only small change in velocity.
Within an LJJ the fluxon is protected from external perturbations
due to its topological nature
(a very large energy is required to change the phase windings of the JJs).
All reversible gates will be powered by the incoming fluxon energy alone and the fluxon and antifluxon are chosen to represent bit states 0 and 1, respectively.

\begin{figure}[pbt]\centering
\includegraphics[width=\textwidth]{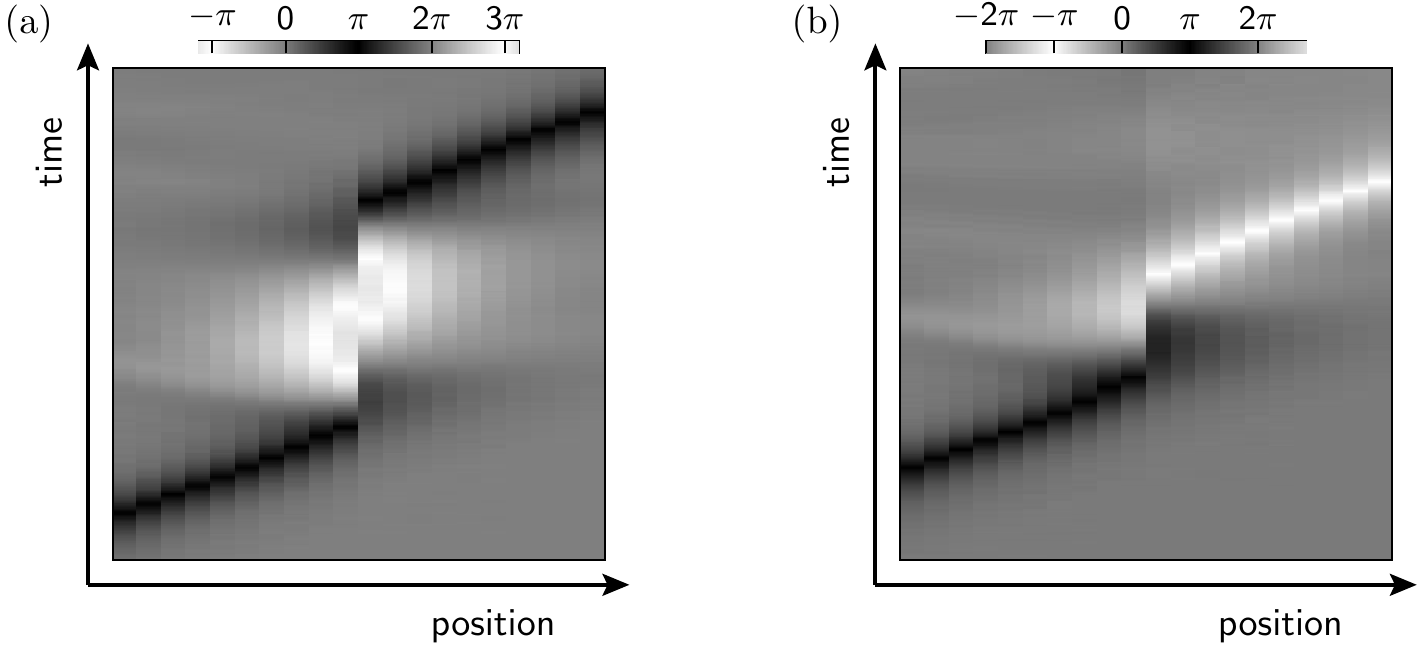}
\caption{
The phases $\phi_i$ of the junctions simulated in 1-bit gates:
(a) Identity (ID) and (b) NOT. At the earliest time the phases are all set to zero,
and in the center position is the interface of the gate.
In simulations a fluxon approaches the interface and exits as a (a) fluxon or (b) antifluxon,
after resonant dynamics enabled by the interface parameters.
Interface parameters of the ID (a) are identical to those in the NOT (b), except $\CJbb \approx 6 C_J$
(see Figure \ref{fig1}).
}
\label{fig2}
\end{figure}

The phase differences $\phi_i$ across each junction $i$, including those labeled $N_l-1$ through $N_l+2$
near the gate interface (dashed box), are dynamical variables of the system.
Phase values of zero (modulo $2\pi$) are potential energy minima.
In Figure \ref{fig2} the phases from (a) an ID gate and (b) a NOT gate are shown in greyscale for a numerical simulation.
In each panel the x-axis represents the positions of 20 JJs along the two gate LJJs
(in sequence) and time is shown along the y-axis.
At the earliest time shown all JJs are in the state $\phi_i=0$ (equivalent phases modulo $2\pi$ operate equivalently).
At a later time the fluxon, whose center with $\phi_i \simeq \pi$ is shown in black in greyscale,
moves to the right, approaching the interface.

In Figure \ref{fig2}(a) we show dynamics for an ID gate, where the parameters are identical to the NOT
(Figure \ref{fig2}(b)), except with a different central capacitance $\CJbb$.
While the ID gate is logically simple, the underlying dynamics is more complex than a direct (ballistic) transmission of fluxon from left to right.
As the fluxon approaches the center position where the interface is located, oscillations start.
The oscillations are powered solely by the incoming fluxon.
Oscillations persist for only a small duration comparable to the natural oscillation (plasma) period of a single JJ in the LJJ (for the duration of a few plasma periods).
As the oscillations stop, a fluxon (black in the greyscale) appears in the right LJJ and moves freely to the right.
This phenomenon is interesting in its own right because it extends soliton
(and fluxon) dynamics beyond previous studies,
including the chaotic fluxon scattering at a perturbation within a LJJ \cite{GooHab2007}.
This is an Identity (ID) gate for our purposes because the input fluxon polarity encoding the bit state is unchanged at the output.
Note that the slope of the black line, in time versus position, is approximately the same before and after the gate.
This indicates that the gate conserves nearly all of the incoming kinetic and potential energy, as all LJJs are the same unless noted otherwise
(potential energy is conserved in the fluxon since the input and output LJJs are equivalent).

Figure \ref{fig2}(b) shows the result for the NOT gate with parameters
described in Figure \ref{fig1}.
The incoming fluxon again approaches the interface,
but different oscillatory dynamics are induced related to a smaller interface capacitance  (see below).
As a result of the altered interface dynamics an antifluxon now exits instead of a fluxon.
The center of this particle appears as white in the greyscale for $\phi_i \simeq -\pi$,
indicating opposite phase winding relative to the fluxon (wrapping in the LJJ from $0$ to $-2\pi$
rather than $0$ to $2\pi$).
In the NOT gate, like the ID, the velocity of the output fluxon is nearly the same as the input fluxon, indicating good energy conservation (and good dynamical reversibility) for the fluxon mode.

These gate dynamics were also analyzed by means of a quasiparticle model,
where one quasiparticle describes the collective many-JJ dynamics on each side of the gate.
This is different in C-SFQ logic where only approximately two JJs (degrees of freedom) might switch per independent motion.
Here the input LJJ has a quasiparticle consisting of a fluxon and mirror antifluxon to represent the incoming fluxon and interface oscillations on that side.
Likewise the output LJJ has an equivalently defined quasiparticle.
With only this collective-coordinate ansatz, the two quasiparticle dynamics for the 1-bit gates is found to agree with the numerical simulations.
It turns out that the effective mass of the quasiparticles changes during the gate operation and the interface capacitance $\CJbb$ creates a large effective mass for the quasiparticles and a strong interaction force between them \cite{WusOsb2017}.
For the NOT gate the smaller interaction force from the smaller effective mass results in shorter-oscillation dynamics, in comparison to the ID gate.

Though these gates are defined for parameters (given in Figure \ref{fig1}) and no bias,
one can of course find dynamics in other regimes including a regime where the fluxon is destroyed.
Also, there is a regime where the fluxon will reflect at the interface as if there is a boundary imposed by the interface, a phenomenon which is well known \cite{DeLeonardisETAL1979}.
In contrast, the acceptable (useful for our purpose) gates for our application have sufficient output velocity to allow the fluxons to quickly reach a subsequent gate.
Parameter margins for gates appear achievable, generally with an acceptable range of over $10 \%$ (not shown).

\section{Two-input RFL gates including the IDSN}

\begin{figure}[ptb]\centering
 \includegraphics[width=\textwidth]{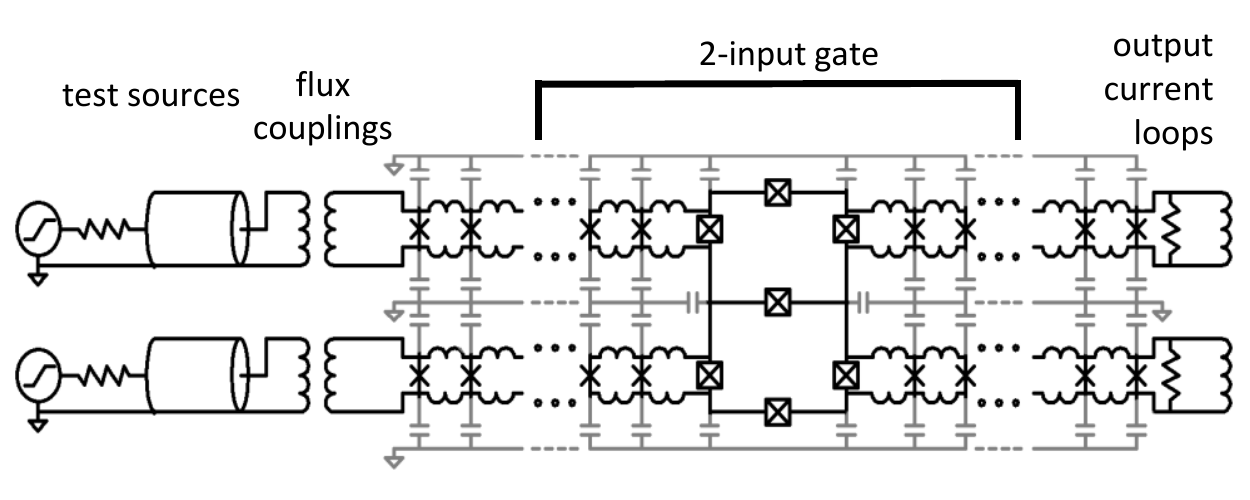}
 \caption{
 A 2-input gate, which may be either an NSWAP or IDSN, in a test structure.
 The gate consists of an interface with 7 JJs (denoted with boxed X's) as well as input and output LJJs.
 Input fluxons approach the gate from the two left (input) LJJs, and output fluxons exit through the two right LJJs, where each LJJ is made of JJs (denoted with unboxed X's).
 Similar to the NOT and ID gate, the capacitances of the central interface JJs can enable oscillatory gate dynamics.
 The number of input fluxons will equal the number of output fluxons
 along paths from left to right.
 Circuit elements to the left and right of the gate are used to simulate these gates in a realistic test environment.
 A capacitively coupled ground is shown in grey wires.
 }
 \label{fig3}
\end{figure}

Figure \ref{fig3} shows a 2-input gate with a proposed test structure which could be implemented experimentally.
A 1-bit gate could be tested in a similar structure.
Here a voltage step at each source can create a data fluxon (or antifluxon) in the input LJJs and induce it to approach the gate with sufficient velocity.
After the gate operation, fluxons exiting the gate approach superconducting loops to the right.
These loops store currents with a direction indicating the output fluxon polarity.
Simulations (not shown) have been performed on this structure and simpler environments, and indicate that the gates are enabled by balanced signals on the LJJs in the gates.
Here the balanced LJJ fluxon signal carries a pulse of voltages and current of equal magnitude and opposite sign along the two terminals of each JJ.
This subtle yet important point also distinguishes our logic from established SFQ logic.
However, it shares this feature with many superconducting qubits.

The central part of Figure \ref{fig3} shows a 2-input gate schematic with two LJJs attached to the left and right of an interface including 7 JJs (denoted with boxed X's).
With specific parameters it can be used to implement the NSWAP and IDSN gates.
The gate allows oscillatory dynamics like that of the 1-bit gates, but now the oscillations can be dependent on two-input fluxon interactions.
There is top-bottom and left-right symmetry in the gate structure such that there are only three unique JJ parameters for the 7 interface JJs (with finite critical current and capacitance, but no added damping).
Similar to the NOT and ID gates, the capacitances of the central interface JJs are generally larger than the ordinary JJ cells of the LJJ to induce strong interactions between the LJJs.
Related to the balanced nature of the fluxon signals, a ground plane shown in grey circuitry allows balanced coupling to the ground, and may represent stray capacitance.
We find in simulations that adding $10\%$ extra (stray) capacitance from each ordinary JJ in the LJJs to this ground plane had no significant effect on the gate operation (relative to omitting it).

\begin{figure}[pbt]\centering
 \includegraphics[width=9.5cm]{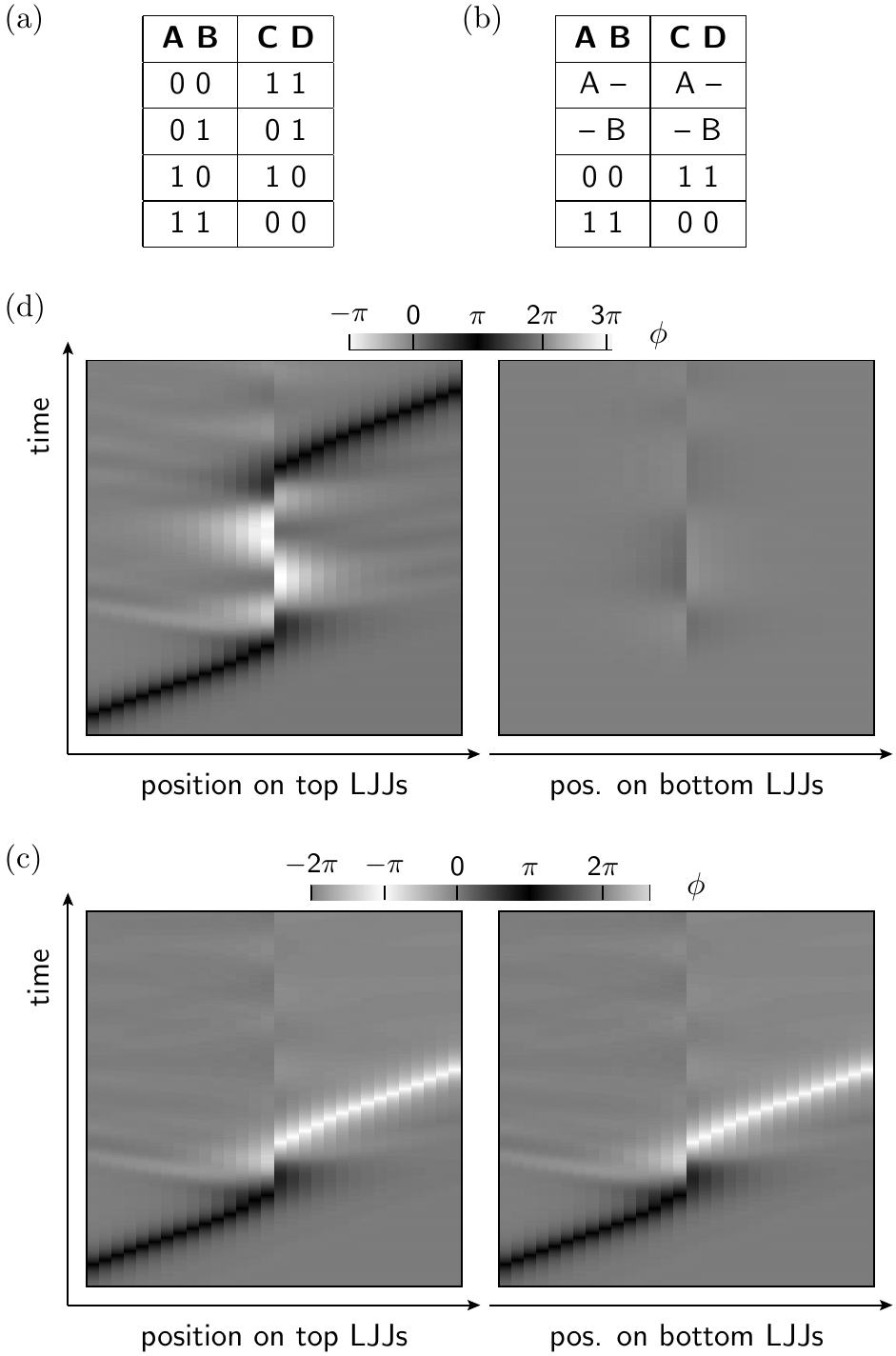}
 \caption{
 Truth table for (a) NSWAP=NOT(SWAP) and (b) IDSN gates.
 The NSWAP gate is reversible for any input combination of fluxon (0) and antifluxon (1).
 The IDSN is defined for any single input fluxon (0 or 1) and fluxons with the same polarity (0 0) or (1 1).
 (c): IDSN gate dynamics for one fluxon into top input LJJ.
 Here an output fluxon only appears at the top output LJJ.
 (d): IDSN gate dynamics for two fluxons (0 0) in the input LJJs.
 In this case two antifluxons (1 1) exit through the output LJJs.
 }
 \label{fig4}
\end{figure}

The logic operation of the NSWAP and IDSN gates are shown in Figure \ref{fig4}(a) and (b), respectively.
The NSWAP is logically equivalent to the common SWAP followed by a NOT of both outputs.
Simulations showed it is robust for computing and it has larger parameter margins than the SWAP gate.
The NSWAP is defined for two input fluxons.
All four of these combinations are efficient in simulation \cite{WusOsb2017}.
A key gate in this work is the IDSN, which is defined to execute a 1-fluxon input as an ID gate,
and a 2-fluxon input of Same polarity as a NOT gate.
Similar to the NSWAP it has good efficiency with reasonable parameter margins.
The gate does not have favorable dynamics when the input polarities are different
(related to having only two favorable projectile paths in a collective coordinate picture).
Therefore we exclude this operation, and only include 2-fluxon input states which fulfill A=B in the IDSN gate definition in Figure \ref{fig4}(b).
These operations are conservative in the number of fluxons (input and output) but interestingly not in the number of 0 and 1 states (the Hamming weight).
This is allowed because gates conditionally change topological charge of individual solitons (fluxons are topological particles).

Figure \ref{fig4}(c) and (d) describe the dynamics of all the logical operations of the IDSN listed in panel (b) because antifluxon inputs undergo the same dynamics up to a symmetry operation (in the case of no fluxon input, of course no dynamics takes place).
In Figure \ref{fig4}(c) the dynamics are shown for 1-fluxon input to the IDSN gate.
This can be seen on the right panel (lower input and output LJJs) as no fluxon input or output but only small phase fluctuations.
Meanwhile the left panel shows the phases in the top input and output LJJs.
At first a fluxon can be seen moving in the top input LJJ and later a fluxon exits at the top output LJJ (right half of x-axis) corresponding to an ID operation.
Again, the nearly unchanged velocity (slope) indicates good energy conservation.
The IDSN is shown for two input fluxons of same polarity in Figure \ref{fig4}(d).
For this case an efficient NOT gate occurs simultaneously across the top LJJs and bottom LJJs, i.e., from inputs to outputs.
While most of the energy from the input fluxon is returned to the output fluxons
(similar to the ID and NOT gate), there is some energy left behind as noise.
However our gate will operate with perfect fidelity because the purpose of the gate is only to transmit the state forward to the next gate, with a minimum specified velocity.

\FloatBarrier
\section{The Controlled-NOT (CNOT) gate}

\begin{figure}[ptb]\centering
\includegraphics[width=9cm, trim=30 32 20 22, clip]{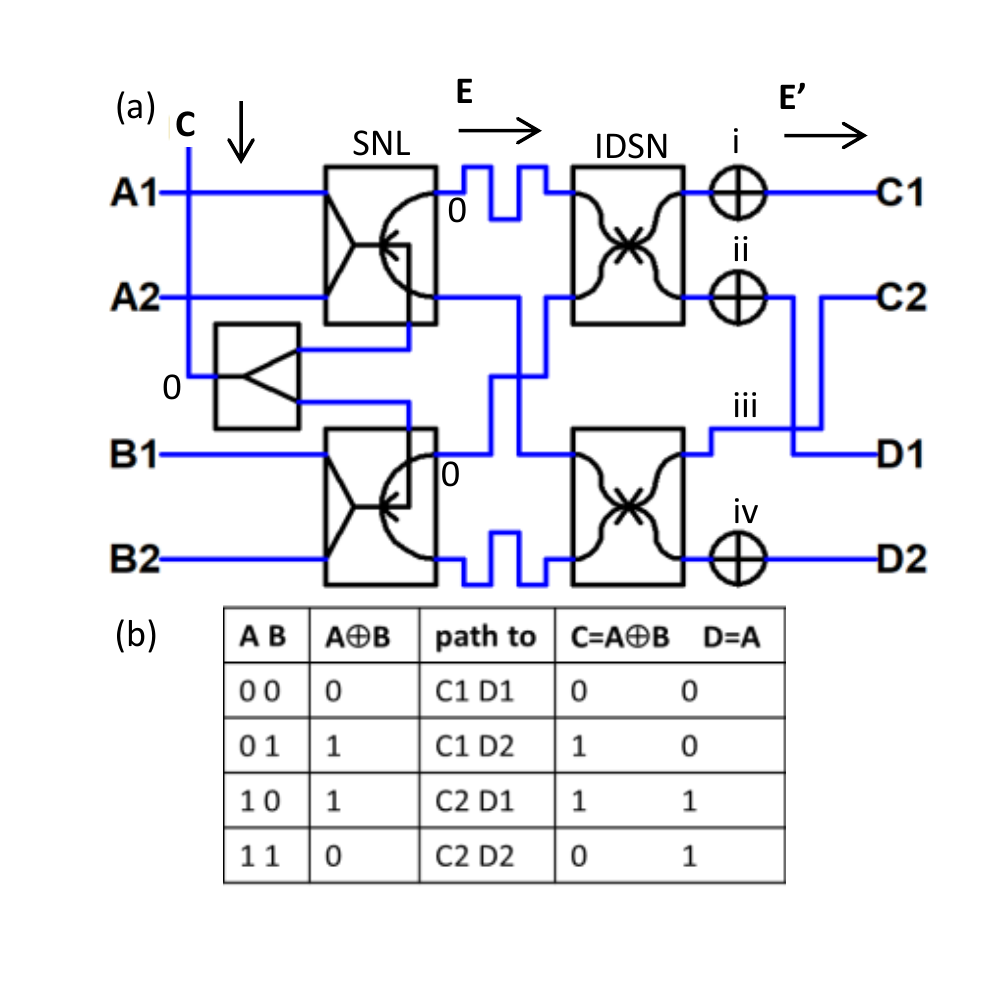}
\caption{
(a): CNOT gate with bit storage schematic where blue lines denote connecting LJJs
and (b) Corresponding fluxon routing and logic table.
 A data fluxon arrives on A1 or A2 and another at B1 or B2.
Each pair of inputs connect to a Store and Launch (SNL) cell to the right.
The bit states of the incoming data fluxons are stored in the SNL cells.
A clock fluxon enters at C and splits into two half-mass fluxons at a T-splitter
(shown as element with ``Y'' shape).
These two lighter clock fluxons synchronously enter the SNL cells and each launch one data fluxon along a bit-dependent path to the right.
The latter fluxons have energy which is much greater than the clock fluxon incident on the SNL.
If the stored bit state is 0 (alternatively 1),
the data fluxon will be launched out the right terminal labeled with 0 (alternatively unlabeled).
The launched data fluxons then enter IDSN gates and some fluxons will experience an additional NOT gate (shown as a circled plus symbol).
The output ports for data fluxons are shown in panel (b) with bit state C (D),
arriving from a fluxon on C1 or C2 (D1 or D2). C carries the CNOT result and D is the copy of A.
The fluxon energy at launch  is similar to the energy  after the reversible gates,
where all free fluxons have potential energy U.
The clock fluxons have a small energy relative to U and enable an efficient digital CNOT (Feynman).
}
\label{fig5}
\end{figure}

Figure \ref{fig5} shows a schematic for the digital CNOT (or Feynman) gate.
The operations take two data fluxons, one through an A (A1 or A2) and one through a B (B1 or B2) input LJJ.
The bit states are stored in Store and Launch (SNL) gates.
The SNL is meant to store most of the energy of the data fluxon, using phase winding of the data LJJs and therefore it is not ballistically reversible like the IDSN and NOT gates described above. 
Afterwards, a clock fluxon arrives along C with a given polarity and is split into two half-energy fluxons
by a T-branch connecting different LJJ types. This process is ballistic and will result in two identical fluxons with the same velocity as the original clock fluxon.
Incidentally, powered T-branches have been previously studied for logic gates
\cite{NakajimaETAL1980}. In our split fluxons, the $L_J$ of the split LJJs is halved with respect to the original for half-energy fluxons. This can be easily understood in terms of the Sine-Gordon boundary conditions \cite{DeLeonardisETAL1979}.
Each of the two resulting clock fluxons then enters its respective SNL to launch a stored fluxon.
The clock fluxons (assumed here to be fluxons with positive polarity), provide the necessary fluxon reformation and kinetic energy.
The process uses static stored energy without bias (ac or dc).
Stored data will be launched as a fluxon along a bit-dependent output LJJ:
an output LJJ marked in Figure \ref{fig5} with 0 is the output path for a fluxon,
while the other output LJJ is made for the antifluxon (1 state).

This CNOT uses three NOT gates at locations marked by \textsf{i}, \textsf{ii}, and \textsf{iv}.
A simpler NCNOT=NOT(CNOT) can be developed by using a NOT only at position \textsf{iii}.
The CNOT and NCNOT have similar utility, and are constructed with two crossovers,
excluding the clock fluxon line.
Excluding these simplest structures, the CNOT and NCNOT gates are made from two SNLs and two IDSNs; more complex 3-input gates can be developed similarly.

Connections between this gate stage and another one will allow gates to be executed in sequence.
For example, two CNOT gates can be cascaded.
Here the outputs $\textrm{C}_k$ ($\textrm{D}_k$) of the first gate with $k=1$ can be connected to the inputs $\textrm{B}_k$ ($\textrm{A}_k$) of the second gate with $k=2$.
With these connections, the second stage output $\textrm{C}$ will be identical to the first stage input $\textrm{B}$ as the second gate uncomputes the first, a feature often created by two-sequential reversible gates.
\section{The Store aNd Launch (SNL) gate}

\begin{figure}[pbt]\centering
\includegraphics[width=\textwidth]{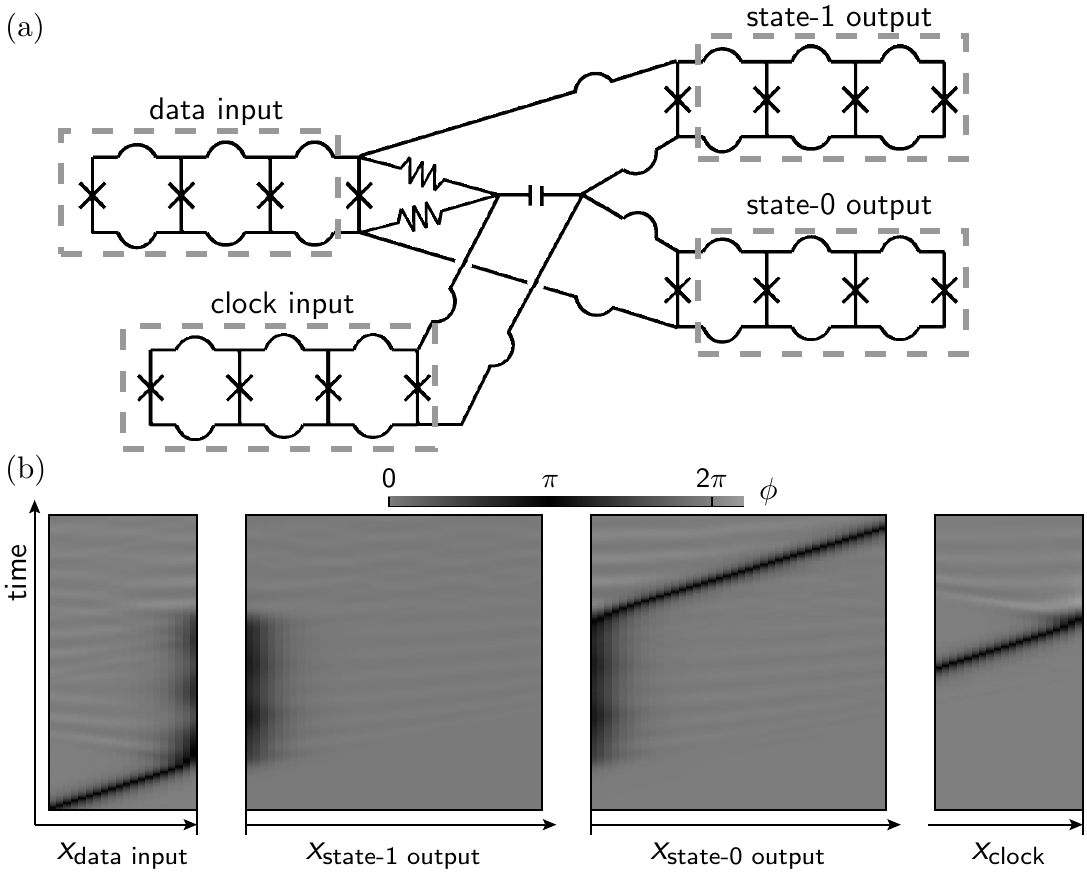}
\caption{
(a) A Store and Launch (SNL) gate schematic containing a data-input LJJ, two launch
LJJs for bi-directional launch and a clock (input) LJJ. The data input and outputs are made from
standard LJJs (in dashed boxes), but the JJs adjacent to the interface cell are designed for high
energy storage.
(b) Numerical simulation results of the schematic.
(b, left panel): The data fluxon (position indicated as black for phase $\phi_i \simeq \pi$)
enters and stops at the interface (shown as tick mark).
For some duration no inputs are given.
(b, right panel): About halfway up the time-scale a clock fluxon enters,
which has $1/4^{\text{th}}$ of the data fluxon energy (carried on a LJJ
with different parameters than the data LJJs).
(b, center panels): When the clock fluxon arrives at the interface it imparts kinetic energy to the stored state and launches it as data fluxon along a polarity-dependent path.
Note that the 0-output path is here shown as the bottom output path while in Figure \ref{fig5} it is the top output path.
}
\label{fig6}
\end{figure}

The SNL gate contains an interface between one (shown in Figure \ref{fig6}) or two (shown in Figure \ref{fig5}) data input LJJ and two data output LJJs.
A clock LJJ is connected to the interface between these LJJs in a symmetric way through two resistors and a capacitor.
The resistors are used to stop an incoming fluxon, and the capacitors have values selected to efficiently transfer the incident clock fluxon energy to the rest of the circuit.
This circuit has enough damping (within an engineered potential) to change an incoming data fluxon into a stored (CW or CCW) circulating current.
It is interesting to note that a previously studied INHIBIT gate controls the routing of a SFQ in a related way to the SNL gate, but it has many differences too: resistive connections to ground,
less symmetry, and the bit state is not defined by flux polarity \cite{OnomiETAL1995}.

As mentioned above for Figure \ref{fig5}, the clock fluxon entering the SNL may already be the result of splitting another clock fluxon in two. Here, with an efficient CNOT gate in mind, the clock fluxon incident on the SNL uses a clock fluxon with $1/4^{\text{th}}$ of the energy compared to a data fluxon. The velocity of the clock fluxon is identical to the data fluxon, and both fluxon types have 80$\%$ of their total energy in potential (or rest) energy.

Using the simulations discussed above as guidance,
the final fluxon energy $E'$ can be made to reach $95\%$ of the input fluxon energy $E$.
Also, the SNL can store approximately the potential energy of the data fluxon $U=0.8E$
(though in the simulation below we have used slightly more favorable energy storage).
The clock fluxon with energy $E/4$ restores and launches a data fluxon after entering the SNL.
Since the activation energy of the data fluxon comes from the clock fluxon and the data fluxons are defined to reach the storage location,
the clock fluxon powers the CNOT, and thus is used for an efficiency estimate below.

Simulations on the SNL are shown in Figure \ref{fig6}(b).
In the leftmost panel the input data fluxon trajectory (black in greyscale) is seen approaching the gate interface (at the rightmost position).
Then black is also seen on the interface, indicating a static circulating current at the LJJ end.
At the same time, phase values $|\phi_i| \simeq \pi$ (black in greyscale)
are seen at the left edge of the two center panels, indicating currents at the output LJJs.
Although only one case is shown, the direction of the circulating current and the sign of these phases would depend on the input-fluxon polarity.
During the storage time a clock fluxon can be seen approaching the gate interface in the right-most panel.
As mentioned before, it has only $1/4^{\text{th}}$ of a data fluxon's energy to create (ballistic) dynamics from the statically stored bit.
When this fluxon reaches the interface the data fluxon is seen as exiting the gate on one of the two center panels.
The output path would be switched if the stored bit had corresponded to opposite circulation as there is symmetry with respect to the two output LJJ paths.
Notice that the output velocity indicated by the slope in the output line is the same as the input line.

An advantage of this combined-use gate is that there is only one position for clock input and dissipation,
unlike C-SFQ which uses dissipative JJ switches throughout the logic cells.
Furthermore, JJ dynamics in the SNL are collective and unbiased such that we expect this gate to be efficient relative to a C-SFQ gate.
As mentioned above, the CNOT is designed such that the entire gate will operate with the energy of the clock fluxon.
The characteristic potential barrier separating the in- and output states of the CNOT is therefore overcome
by expending the small energy of the clock fluxon.
If the potential barrier is set by the requirement of thermal stability, one can estimate
that the clock fluxon might only require and energy $E_{\text{clock fl.}}= \gamma_T k T$.
This saves a factor of $4\pi n$ compared with the typical dissipation in a C-SFQ gate,
as discussed in Section 1.
Based on these estimate, the clock fluxon might only require an energy on the same order of magnitude as the meta-stability energy $E_{\text{clock fl.}} \sim \gamma_T k T$.
Adding memory to nearly ideal-efficiency reversible gates (mainly the IDSN) for the CNOT makes the dominant dissipation cost of the former seem worthwhile at the time of this writing.
The entire CNOT with included memory is expected to attain REEF.
The above inputs allow us to estimate a 1--2 order-of-magnitude energy savings relative to a 2-bit C-SFQ gate.

\FloatBarrier
\section{Conclusion}

In conclusion, we have shown schematics and simulations of ballistic Reversible Fluxon Logic (RFL) gates.
The fundamental RFL gates have no dissipation and are synchronous.
However, to make them more useful, we combine them with bit storage in this work.
Two fundamental gates in RFL are the NOT and ID (Identity) gates and use polarity as bit states.
A new RFL gate primitive named IDSN executes a one-fluxon input as an Identity (ID), and the Same 2-bit-input as a NOT.
A CNOT gate schematic is also shown, a gate not currently available to SFQ logic despite its importance in reversible digital logic.
Our CNOT gate uses a bit Storage aNd fluxon Launch (SNL) gate which contains bit memory.
Two SNL, and even the entire CNOT, are powered by clock fluxon entering the gate.
In one SNL simulation, potential energy is stored from a data fluxon and then a clock fluxon
with only $1/4^{\text{th}}$ of the energy of the data fluxon is shown to launch the stored bit as data fluxons with a velocity nearly equal to its input velocity.
This gate also launches the data (output bit) in one of two paths, dependent on the initial data fluxon (input bit).
This bit-dependent routing from two synchronized SNL gates provides suitable input states for two IDSN gates (gate primitives) within the CNOT.
This work shows an example of how ballistic SFQ logic gates, with bit storage gates for clocking, can result in a gate with reversibly-enabled energy efficiency (REEF).
A 1-2 order-of-magnitude energy savings relative to a 2-bit C-SFQ gate is estimated.
\\~\\

\noindent Acknowledgements:
The authors generally, and the first author specifically, would like to thank Q. Herr, V. Yakovenko, V. Manucharyan, S. Holmes, and M. Frank for helpful discussions during the writing of this manuscript.

%

\end{document}